\documentclass[11pt,a4paper]{article}
\usepackage[round,authoryear]{natbib}
\usepackage[T1]{fontenc}
\usepackage[utf8]{inputenc}
\usepackage[a4paper,margin=0.7in]{geometry}
\usepackage{graphicx}
\usepackage{multicol,multirow}
\usepackage{amsmath,amssymb,amsfonts}
\usepackage{mathrsfs}
\usepackage{amsthm}
\usepackage{algorithm}
\usepackage{algpseudocode}

\usepackage{subcaption}
\usepackage{rotating}
\usepackage{wrapfig}

\usepackage{appendix}
\usepackage{ifpdf}
\usepackage[T1]{fontenc}
\usepackage{newtxtext}
\usepackage{newtxmath}
\usepackage{textcomp}
\usepackage{xcolor}
\usepackage{lipsum}
\usepackage[colorlinks,allcolors=blue]{hyperref}

\theoremstyle{definition}

\numberwithin{equation}{section}

\title{\textbf{Leveraging Scale Separation and Stochastic Closure for Data-Driven Prediction of Chaotic Dynamics}}

\author{
Ismaël Zighed$^{1,2,3}$\thanks{Email: ismael.zighed@sorbonne-universite.fr} \and
Nicolas Thome$^{2,4}$ \and
Patrick Gallinari$^{2}$ \and
Taraneh Sayadi$^{5}$\\[0.5em]
\small $^{1}$Institut Jean le Rond d’Alembert, Sorbonne Université, Paris, France\\
\small $^{2}$ISIR, Sorbonne Université, Paris, France\\
\small $^{3}$SCAI, Sorbonne Université, Paris, France\\
\small $^{4}$Institut Universitaire de France, Paris, France\\
\small $^{5}$M2N, Conservatoire National des Arts et Métiers, Paris, France
}

\date{\small \today}

\newcommand{\keywords}[1]{\vspace{1em}\noindent\textbf{Keywords: }#1}

\begin{document}

\maketitle

\keywords{Reduced Order Modelling, Dynamical systems, Turbulence, Probabilistic modelling, Stochastic modelling, Data-driven closure for turbulence modelling}

\begin{abstract}

Simulating turbulent fluid flows is a computationally prohibitive task, as it requires the resolution of fine-scale structures and the capture of complex nonlinear interactions across multiple scales. This is particularly the case in direct numerical simulation (DNS) applied to real-world turbulent applications. Consequently, extensive research has focused on analysing turbulent flows from a data-driven perspective. However, due to the complex and chaotic nature of these systems, traditional models often become unstable as they accumulate errors through autoregression, severely degrading even short-term predictions. To overcome these limitations, we propose a purely stochastic approach that separately addresses the evolution of large-scale coherent structures and the closure of high-fidelity statistical data. \\
To this end, the dynamics of the filtered data (i.e. coherent motion) are learnt using an autoregressive model. This combines a VAE and Transformer architecture. The VAE projection is probabilistic, ensuring consistency between the model’s stochasticity and the flow’s statistical properties. The mean realisation of stochastically sampled trajectories from our model shows relative $L_1$ and $L_2$ distances of 6\% and 10\%, respectively, with test set.
Moreover, our framework enables the construction of meaningful confidence intervals, achieving a Prediction Interval Coverage Probability (PICP) of 80\% with minimal interval width. \\
To recover high-fidelity velocity fields from the filtered latent space, Gaussian Process (GP) regression is employed. This strategy has been tested in the context of a Kolmogorov flow exhibiting chaotic behaviour analogous to real-world turbulence. We compare the performance of our model with state-of-the-art probabilistic baselines, including a VAE and a diffusion model. We demonstrate that our Gaussian process-based closure outperforms these baselines in capturing first and second moment statistics in this particular test bed, providing robust and adaptive confidence intervals. When using \(L_1\) and \(L_2\) norms to evaluate the first moment, our GP surpasses its closest competitor by 49\% and 48\%, respectively. For second-moment statistics, assessed via the Continuous ranked Probability Score (CRPS), the GP outperforms both baselines by nearly 100\%.
\end{abstract}

\section{Introduction}

When modelling turbulence, the deterministic toolbox encounters two notable challenges. The first relates to the chaotic and unpredictable nature of the dynamical system, which exhibits erratic trajectories seemingly devoid of recurrent patterns and is highly sensitive to initial conditions. As Lorenz famously put it, “The present determines the future, but the approximate present does not approximately determine the future”. The second hurdle arises from the problem's complex, multi-scale nature. The various scales involved pose difficulties for both DNS, due to the need for excessive constraint on the resolution to account for all the scales, and for machine learning models, which struggle to represent their inherent complexity. 

From a reduced-order modelling perspective (with ML strategies being a subcategory of this), we believe that decomposing the task into two consecutive subtasks will overcome both of the aforementioned hurdles. The \textbf{``first task"} would be to derive the dynamics from low-pass filtered data, focusing on the evolution of large structures that contribute significantly to the total kinetic energy. This approach is commonly taken in the fluid mechanics community by relying on Large Eddy Simulations (LES) (see, for example, \citep{LES_originalPaper}) as opposed to Direct Numerical Simulation (DNS). However, as with LES, the question of closure remains. In the \textbf{``second task"}, therefore, once the trajectories have been reconstructed in the filtered space, we hypothesise the existence of a time-independent and stochastic mapping from that space to the full-resolution domain.

The first task involves constructing an autoregressive and predictive Reduced-Order Model (ROM).  The Reduced Order Model (ROM) framework has demonstrated promising capabilities across a wide range of dynamical systems, including chaotic ones \citep{ROMFluidFlows}. For instance, Proper Orthogonal Decomposition (POD) enables the projection of a system’s states onto lower-dimensional spaces, which are spanned by the system’s space-time coherent structures referred to as "modes." As an analytically derived technique, POD uncovers interpretable modes and has been widely adopted as a powerful baseline for dimensionality reduction in physics and turbulence modelling \citep{POD_Schmid}. In the nonlinear dynamics community, there has been growing interest in Koopman theory, which supports the existence of linear dynamics even for strongly non-linear systems through the infinite-dimensional Koopman operator \citep{Koopman}. Many works and algorithms focus on finding tractable, finite-dimensional approximations of this operator, that capture the essential dynamics. Among these, Dynamic Mode Decomposition (DMD) is one of the most prominent examples \citep{DMD}. Another powerful approach involves identifying the slowest invariant manifold spanned by the eigenspace of the linearized system near a fixed point and extending it through the system’s nonlinearities. This method, known as Spectral Submanifolds (SSMs), reveals low-dimensional, non-linear structures that allow for model reduction and simplification of the dynamics \citep{SSM_1}. Alternatively, these low-dimensional structures can also be identified using machine learning (ML), thereby relaxing the linear or orthogonality constraints inherent to POD. These ROM approaches can generally be divided into two main categories: Intrusive, which are tailored to the Navier–Stokes equations via for example Galerkin projection \citep{CD-ROM}, and non-intrusive, which are purely data-driven \citep{ILED,ESN,MZ,VpROM}. Some non-intrusive approaches even operate in a mesh-independent fashion \citep{infinity}. Recent advances in embedded memory architectures, particularly attention mechanisms and transformers \citep{AttentionAllYouNeed} for time-series applications, have considerably reshaped this landscape \citep{SLT,AutoencoderTransformer,UPdROM}. 

Regardless of the strategy employed, ROMs typically rely on some form of dimensionality reduction, such as proper orthogonal decomposition (POD), autoencoders or other compression techniques. This is based on the assumption that the system’s dynamics evolve on a low-dimensional manifold in phase space \citep{SSM_1,SSM_2}. Remarkably, this assumption often holds for chaotic systems \citep{ChaoticSSM}, with the infamous Lorenz butterfly serving as a paradigmatic example of such low-dimensional underlying structure. However identifying the attractor is not straightforward, and many attempts have shown that, due to inherent turbulence, the resulting models tends to diverge from the true trajectory used in forecast mode \citep{ChaoticSSM,ozalp}.

Therefore, in this paper, we take a stochastic approach to overcome this challenge. To this end, we use a probabilistic reduction strategy involving an encode-process-decode architecture with a transformer in the latent space and a VAE to generate stochasticity. This enables us to learn the time evolution of the system in the filtered space (i.e. large-scale dynamics) and generate a stochastic ensemble of trajectories whose statistical properties closely resemble those of the filtered flow. A second probabilistic model is then employed to close the dynamics for smaller eddies, generating new samples that are statistically consistent with the true distribution (DNS data) in both time and space. This "closure" is related to the core idea of super-resolution, whereby high-resolution data or fields are reconstructed from coarser measurements or simulations~\citep{SuperRes_CNN}. 

Similar super-resolution approaches have been successfully applied in many different fields, including image processing. \citep{SuperRes_Image_1, SuperRes_Image_3}, climate modelling \citep{SuperRes_Climate_1,SuperRes_Climate_2}, fluid dynamics \citep{SuperRes_Flow_1,SuperRes_Flow_2}, and medical imaging \citep{SuperRes_Medical_1,SuperRes_Medical_2,SuperRes_Medical_3}, where capturing detailed spatial or temporal patterns is critical for accurate analysis and prediction. For such small-scale closure, or "super-resolution" task, many studies have utilized groundbreaking advances from the image generation field, such as Generative Adversarial Network (GAN) \citep{GANs} and Variational Autoencoder (VAE) \citep{VAE}. In the 2020s, denoising diffusion models have emerged as a major breakthrough \citep{DDPM,Diffusion_2}. They have enabled state-of-the-art tools for high-resolution image synthesis, although their penetration into physics-based applications remains more limited. Notable examples include diffusion models with embedded transformers. For instance, ~\citep{AROMA}, in their AROMA framework, have successfully leveraged this approach to develop a robust and flexible PDE emulator that adapts to different geometries and scales. ~\citep{Diffusion_closure} integrated a deterministic reduced-order model (ROM) for large-scale dynamics with a diffusion-based closure to improve climatic mesh refinement. Diffusion models, along with the extensive research accompanying them, have demonstrated impressive capabilities in learning complex probability distributions under conditioning; however, they remain computationally expensive both for training and inference \citep{Diffusion_3}. In this context, we propose an alternative approach for probabilistic closure, leveraging Gaussian Processes to achieve an efficient yet expressive representation of small-scale dynamics. 

The paper is organised into three main sections. First, in section \ref{sec:Kolmogorov} we describe the Kolmogorov flow, the data derived from simulating it, and the filtering strategy. Next, in section \ref{sec:ROM} we present the Reduced Order Model (ROM) and its performance on the filtered trajectories. Finally, we introduce the Gaussian Process Closure and detail its theoretical foundations in section \ref{sec:Closure}. We also compare its results against other probabilistic baselines whose implementation details are in the Appendix \ref{subsec:alternative_arch}. Paragraph \ref{sec:conclusion} delivers concluding remarks.

\section{Problem formulation}
\label{sec:Kolmogorov}
\subsection{Kolmogorov Flow}
The test case considered is the \textit{Kolmogorov flow}, a two-dimensional, incompressible, unsteady solution of the Navier–Stokes equations with doubly periodic boundary conditions, $[x,y] \in \mathbb{R}^2 = [0, 2\pi] \times [0, 2\pi]$. It is driven by a sinusoidal forcing term applied in the $x$-direction. The governing equations are:

\begin{equation}
\label{NS}
\begin{aligned}
\frac{\partial \mathbf{u}}{\partial t}
    &= -\nabla p - (\mathbf{u} \cdot \nabla) \mathbf{u} + \frac{1}{Re} \nabla^2 \mathbf{u} + \mathbf{f}, \\
\nabla \cdot \mathbf{u}
    &= 0, \\
\mathbf{f}(x, y)
    &= A \sin(n_f y)\, \mathbf{e}_x
\end{aligned}
\end{equation}

where $\mathbf{u} = (u, v)$ is the velocity field, $p$ the pressure, $\nu$ the kinematic viscosity, $Re$ the Reynolds number and $\mathbf{f}$ the external forcing. The forcing function $\mathbf{f}$ introduces a steady, spatially periodic body force in the $x$-direction with wavenumber $n_f$ and a fixed amplitude $A$.

This flow is well-suited for investigating fluid instability and the transition to turbulence \citep{Kolmogorov}, displaying a variety of regimes depending on the forcing wavenumber $n_f$ and Reynolds number $Re$ \citep{KolmogorovRe_Nf}. For sufficiently high $Re$, an energy cascade is exhibited, whereby large coherent structures decay into smaller eddies that contribute less energy and are eventually dissipated by viscosity. This is the regime of interest to this work. Here, we consider $Re = 34$ and forcing wavenumber $n_f = 4$, inspired by recent studies from ~\citep{kolmoMagri} and ~\citep{racca_2023}. These parameters result in a weakly turbulent flow, where different scales interact chaotically together, motivating their study through a statistical lens.

\begin{figure}[ht]
    \centering
    \begin{subfigure}[t]{0.48\textwidth}
        \centering
        \includegraphics[width=1.2\linewidth]{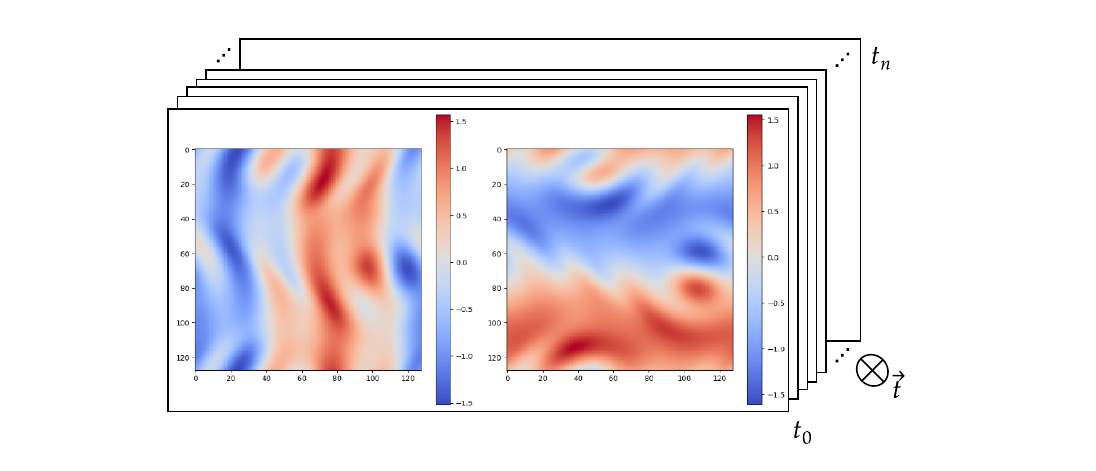}
        \caption{U and V velocity fields}
        \label{fig:UV}
    \end{subfigure}
    \hfill
    \begin{subfigure}[t]{0.48\textwidth}
    \hspace*{-0.1\textwidth}
        \centering
        \includegraphics[width=1.2\linewidth]{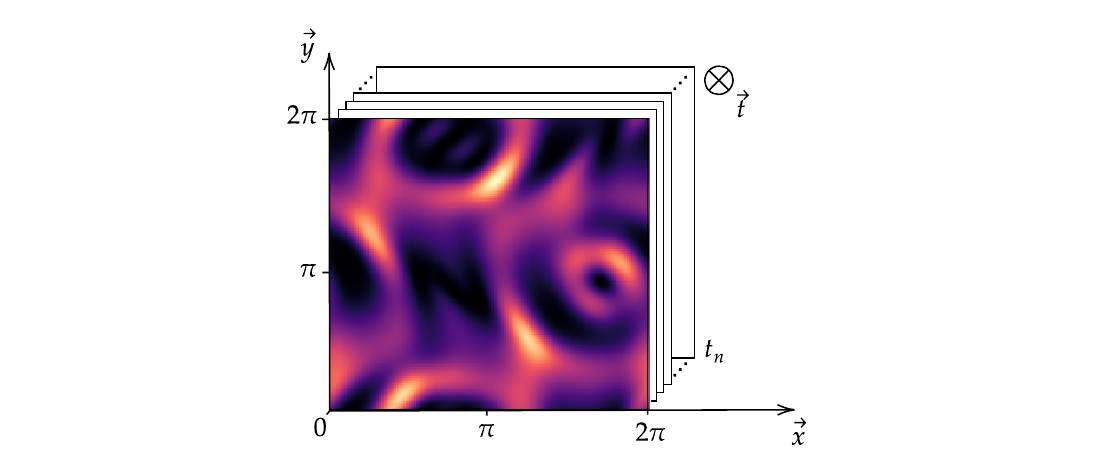}
        \caption{Kinetic energy}
        \label{fig:K}
    \end{subfigure}
    \caption{Velocity fields and corresponding kinetic energy signal.}
    \label{fig:UV_K}
\end{figure}

The simulation is performed on a $64 \times 64$ Eulerian grid using the publicly available pseudospectral solver \textit{kolSol}, based on the Fourier–Galerkin approach described by \citep{Canuto}. We use $64$ Fourier modes for the resolution. The equations are solved in Fourier space using a fourth-order explicit Runge–Kutta integration scheme with a time step of $dt = 0.01$, and results are stored every $\delta t = 0.12$. Velocity fields, shown in figure \ref{fig:UV}, are matrices defined over $\mathbb{R}^{N_x\times N_y \times T}$ with $N_x = N_y = 64$ and $T = 1000$, defining the total number of snapshots. The resulting kinetic energy signal, $K$, is computed by equation \ref{kinetic} and illustrated in figure \ref{fig:K}. 
\begin{equation}\label{kinetic}
        k_t = \frac{1}{2N_xN_y} \sum_{d=1}^{N_xN_y} \left( u_{d,t}^2 + v_{d,t}^2 \right)
\end{equation}

\begin{figure}[H]
    \centering
    \includegraphics[width=0.8\linewidth]{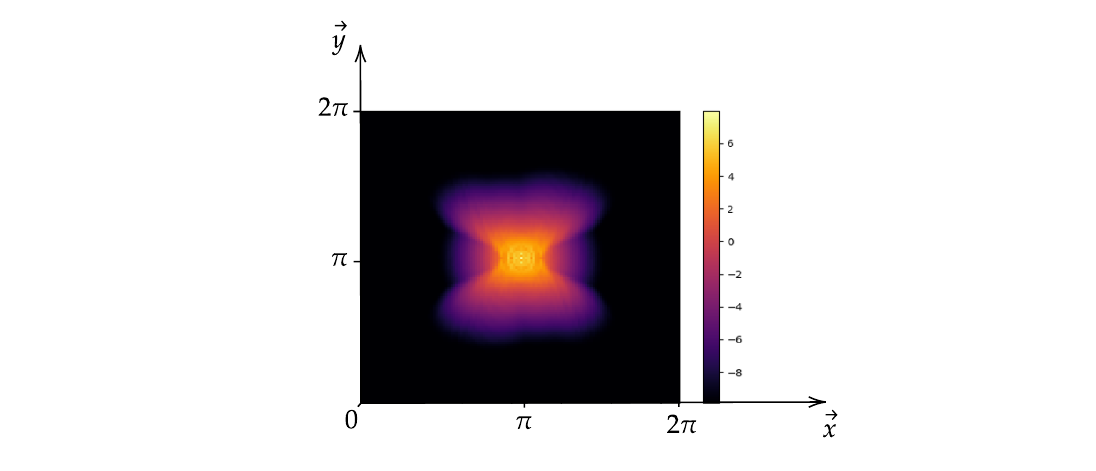}
    \caption{Time-averaged energy spectrum displayed on a logarithmic scale.}
    \label{fig:F(k)}
\end{figure}

The energy cascade, central to Kolmogorov and Richardson's turbulence theories, is based on decomposing the flow's energy into structures of different sizes or, equivalently, into different frequencies, also referred to as wavenumbers. Larger structures correspond to lower wavenumbers and typically carry more energy. Our approach relies on separating these scales under the assumption that there exists a dichotomy between large coherent structures and the smaller eddies into which they break down.
The spatial distribution of these structures, or more precisely their distribution across wavenumbers, is obtained by applying the two-dimensional (2D) Fourier transform to the field, denoted as:
\[
U(x, y) \mapsto \hat{U}(m, n) = \mathcal{F}_{2D}[U](m,n)
\]
where the transform is defined as:
\[
\mathcal{F}(m,n) = \int_{-\infty}^\infty \int_{-\infty}^\infty U(x,y) \, e^{-j 2\pi (m x + n y)} \, dx \, dy.
\]
Here, \( m \) and \( n \) are the spatial frequencies (or wavenumbers) in the \( x \)- and \( y \)-directions, respectively. The quantity \( |\mathcal{F}(m,n)| \) represents the magnitude spectrum of the kinetic energy field \( K \). It provides the distribution of wavenumbers at any given timestep. The time-averaged spectrum is shown in Figure \ref{fig:F(k)} on a logarithmic scale.

The spectrogram indicates that the structures with the highest energy are concentrated near the centre, corresponding to small wavenumbers. As the wavenumber increases, moving away from the center, the magnitude generally decreases, reflecting the diminishing contribution of smaller-scale structures to the kinetic energy. Our computed energy cascade is illustrated in Figure \ref{fig:threshold}.

\subsection{Strategy for scale separation}

\begin{figure}[H]
    \centering
    \includegraphics[width=0.5\linewidth]{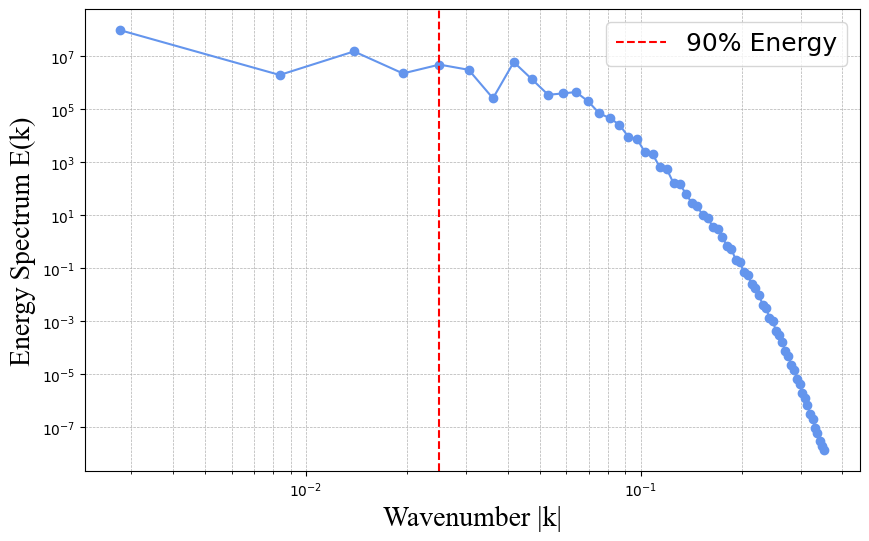}
    \caption{Low-pass filter threshold keeping 90\% of the energy}
    \label{fig:threshold}
\end{figure}

\begin{figure}[H]
    \centering
    \includegraphics[width=0.8\linewidth]{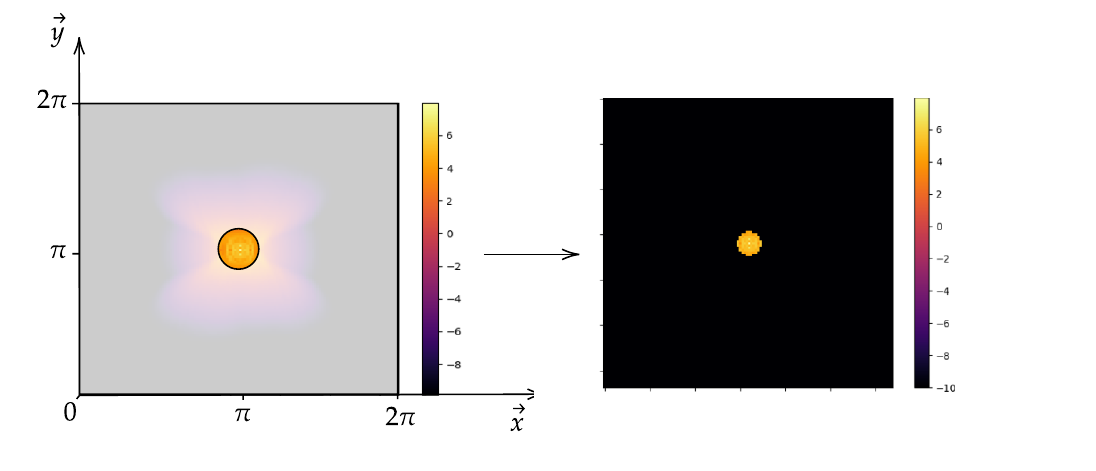}
    \caption{Low-pass filter on the energy spectrum and filtered energy spectrum}
    \label{fig:low-pass}
\end{figure}

\begin{figure}[H]
    \centering
    \begin{subfigure}[t]{0.32\textwidth}
    \hspace*{-0.4\textwidth}
        \centering
        \includegraphics[width=2\linewidth]{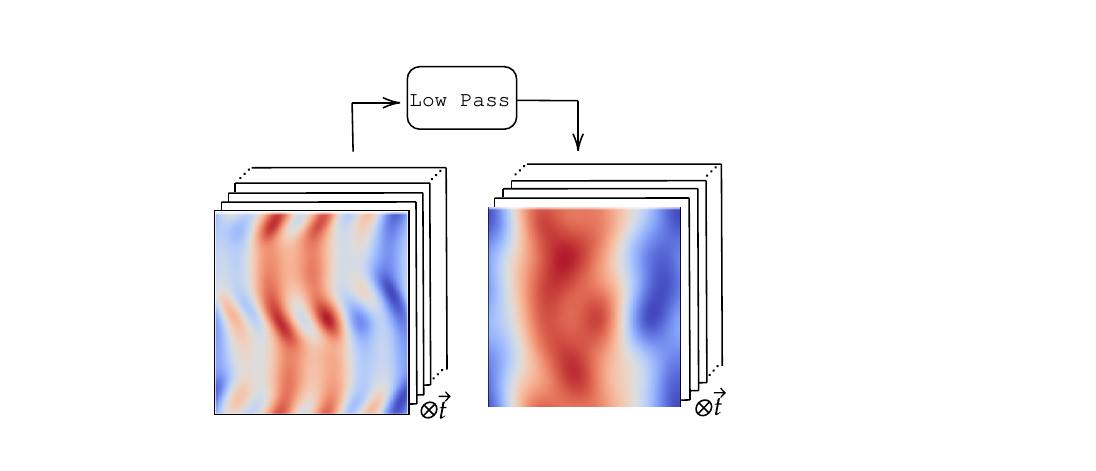}
        \caption{Low-pass filter effect on velocity field U}
        \label{fig:lp_U}
    \end{subfigure}
    \hfill
    \begin{subfigure}[t]{0.32\textwidth}
    \hspace*{-0.4\textwidth}
        \centering
        \includegraphics[width=2\linewidth]{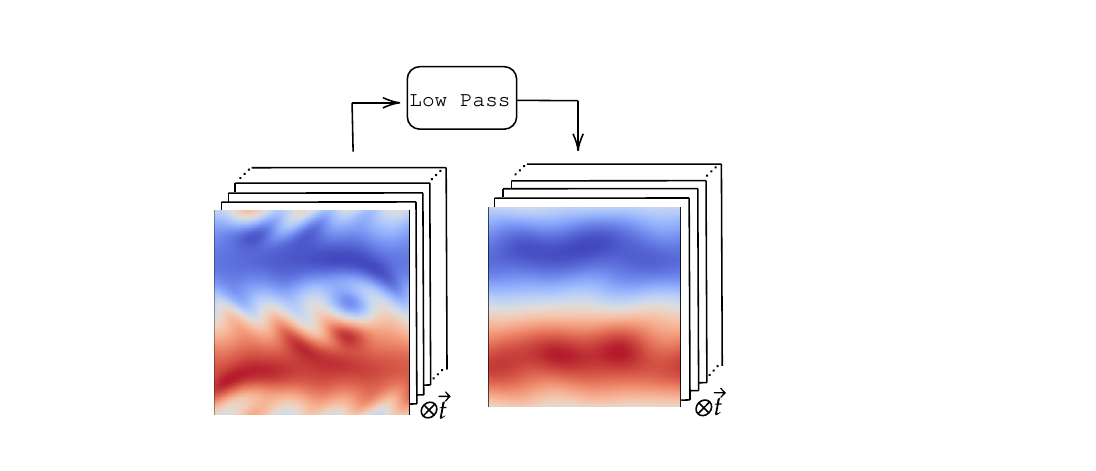}
        \caption{Low-pass filter effect on velocity field V}
        \label{fig:lp_V}
    \end{subfigure}
    \hfill
    \begin{subfigure}[t]{0.32\textwidth}
    \hspace*{-0.4\textwidth}
        \centering
        \includegraphics[width=2\linewidth]{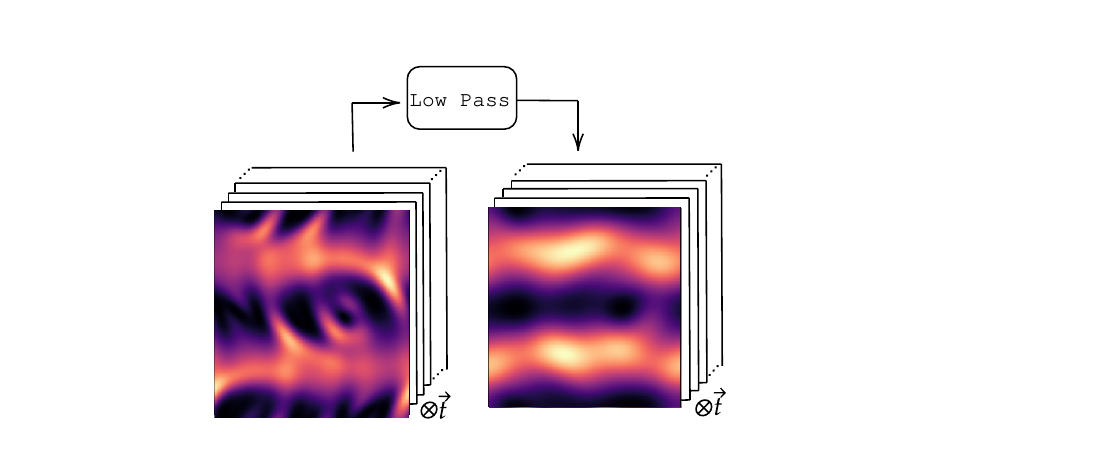}
        \caption{Low-pass filter effect on the kinetic energy K}
        \label{fig:lp_K}
    \end{subfigure}
    \caption{Effect of low-pass filter on velocity fields and kinetic energy.}
    \label{fig:lp_UVK}
\end{figure}

In this work, we propose to classify the flow scales into two families: large and small structures. These two families will be addressed with two distinct strategies. Our goal is to identify an energy threshold in the cascade that retains the larger, "simpler" structures responsible for approximately 90\% of the total kinetic energy. This threshold, illustrated Figure \ref{fig:threshold} corresponds to a critical wavenumber $k_c = 0.03$, which is then used to construct a low-pass filter $H$ that separates the large-scale motions from the smaller-scale fluctuations.

\[
H(m,n) = 
\begin{cases}
1, & \text{if } \sqrt{m^2 + n^2} \leq k_c, \\
0, & \text{otherwise}.
\end{cases}
\]
Applying this filter to the Fourier-transformed field, we obtain the filtered spectrum:
\[
\hat{U}_{\text{low}}(m,n) = H(m,n) \cdot \mathcal{F}(m,n).
\]
Finally, the filtered velocity field \( U_{\text{low}}(x,y) \) in physical space is recovered by the inverse 2D Fourier transform:
\[
U_{\text{low}}(x,y) = \int_{-\infty}^\infty \int_{-\infty}^\infty \hat{U}_{\text{low}}(m,n) \, e^{j 2\pi (m x + n y)} \, dm \, dn.
\]
This operation isolates the large-scale structures corresponding to wavenumbers below \( k_c \), filtering out the smaller-scale fluctuations as illustrated Figure \ref{fig:low-pass}.
The filter reduces the spatial complexity and fine-scale features of the flow that pose challenges for a Reduced Order Model (ROM) to accurately capture \citep{ROM_limitation_POD,ROM_limitations}.
The effects of the filter on the velocity fields and kinetic energy are illustrated in Figures \ref{fig:lp_U}, \ref{fig:lp_V}, and \ref{fig:lp_K}, respectively.

The filtered data resulting from this process will be used to train the dynamic Reduced-Order Model (ROM) described in the next section.  
\section{First task: Predicting Filtered dynamics}
\label{sec:ROM}
\subsection{Model architecture}

We employ a Variational Autoencoder (VAE) to identify a reduced latent space onto which the dynamics are learned by a Transformer, leveraging recent advances in embedded memory and attention mechanisms \citep{AttentionAllYouNeed,SLT}. The Mori-Zwanzig formalism and Takens’ theorem \citep{Taken,Mori,Zwanzig} emphasize the importance of incorporating a memory term to effectively account for the influence of the unobserved, orthogonal subspace on the set of observables learned through the VAE projection. Additionally, \citep{MZ} demonstrate the limitations of using a purely Markovian representation for rolling out dynamics within a finite, invariant Koopman space. Our architecture, illustrated in Figure~\ref{fig:ROM}, is the same used in the framework \textit{UP-dROM} \citep{UPdROM}.

\begin{figure}[H]
\centering
\includegraphics[width=0.9\linewidth]{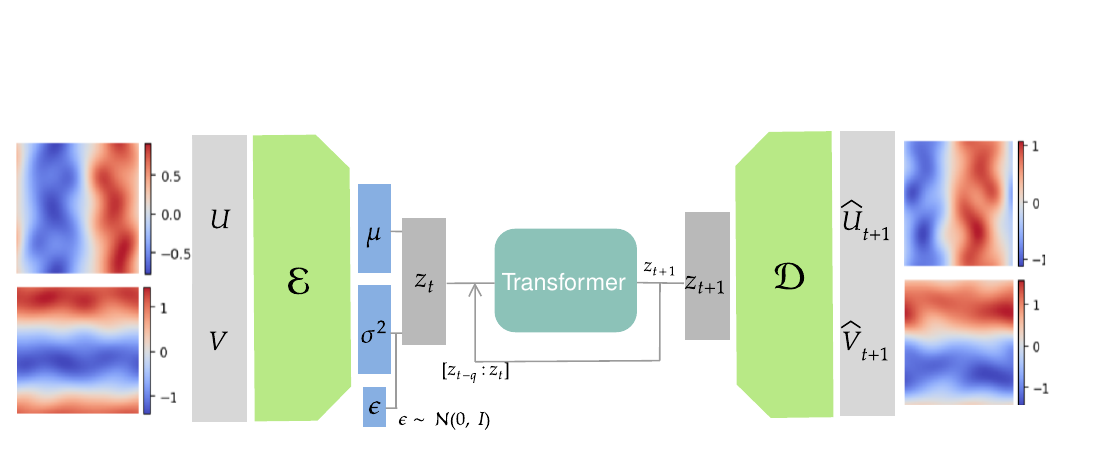}
\caption{Reduced-Order Model architecture, used to predict filtered trajectories.}
\label{fig:ROM}
\end{figure}

This model has demonstrated robust performance and generalisation capabilities in a deterministic setting involving two-dimensional, incompressible flow passed a cylinder. The theoretical foundations and implementation details of that ROM are publicly available on GitHub \citep{GitHubRepo} and in UP-dROM.
Particularly interesting here is the use of a Variational Autoencoder (VAE), since, apart from its ability to produce a well-structured and dense latent space, which enhances generalisation performance, the VAE learns a distribution over latent variables. This enables the generation of diverse yet physically plausible trajectories. This probabilistic framework therefore naturally supports stochastic sampling and uncertainty quantification, as demonstrated in UP-dROM. 

The VAE learns a mapping $f : \mathbb{R}^{64\times64\times 2} \rightarrow \mathbb{R}^{64}$, and the transformer is trained jointly, leveraging self-attention through two attention blocks to learn the temporal dynamics on this latent manifold. Details of the architecture and hyperparameters are provided in Appendix~\ref{subsec:ROMarch}.
To provide the model with sufficient prior knowledge of the latent data distribution, we train it on a set of eight realisations of the flow that share similar properties but differ slightly in their initial conditions. The dataset is split using an odd-even scheme, with half of the snapshots used for training and the other half for testing. Additionally, we use only the first 80\% of the train and test snapshots, and combine the remaining 20\% to assess the model's ability to reproduce long-term statistical behaviour. As a result, both the training and test sets contain 400 snapshots each, while the evaluation set contains 500 snapshots.
Once the model generates predictions beyond the 500 snapshot mark, we compare the predicted flow statistics with the ground truth statistics to evaluate long-term performance. To keep the presentation clear and succinct, we present results for a single flow instance. The model is tasked with reconstructing the flow from its initial condition and must then roll out the dynamics over an extended prediction horizon, while maintaining statistical consistency with the true flow it aims to emulate. We emphasize that the model could also be evaluated from a more \textit{generative} perspective, as it is fully capable of generating new flow instances when provided with previously unseen initial conditions. However, we consider this experimental setup to be beyond the scope of the present study.

\subsection{Model predictions}
\label{ROM-perf}

\begin{figure}[H]
    \begin{minipage}{\linewidth}
        \centering
        \hspace*{-2cm}  
        \includegraphics[width=1.2\linewidth]{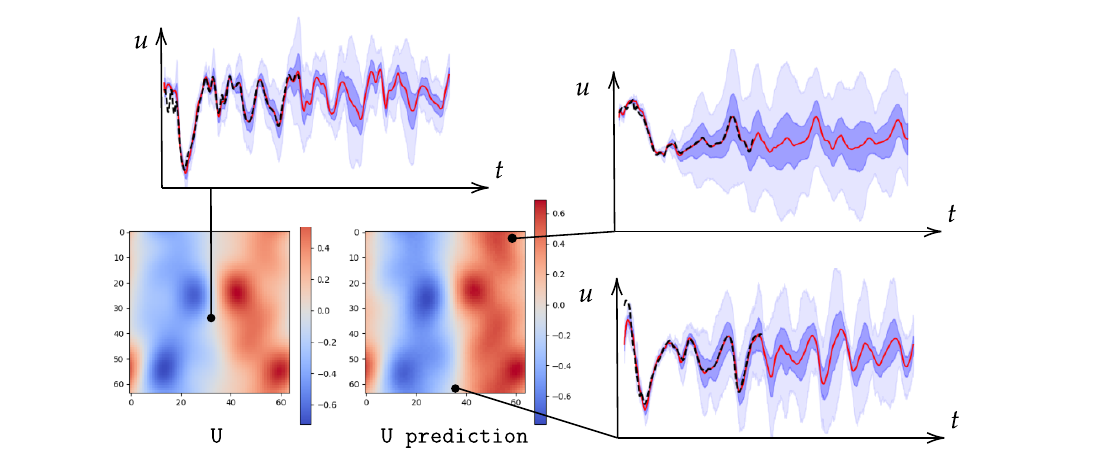}
        \hspace*{-2cm}
        \includegraphics[width=1.2\linewidth]{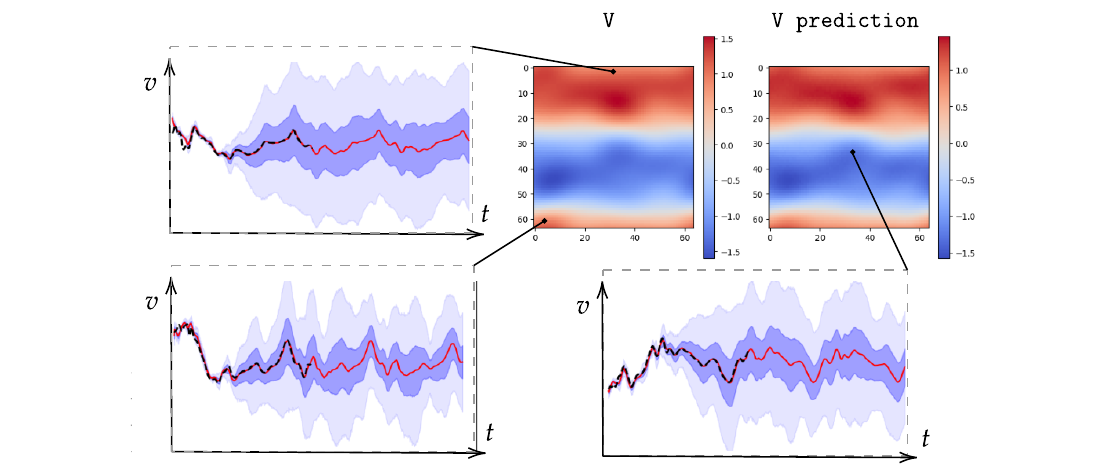}
        \includegraphics[width=0.6\linewidth]{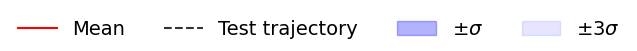}
        \caption{Comparison of predicted and true trajectories for both flow fields at a given time. The plot also shows an ensemble of sampled trajectories generated by the ROM at different locations. The \textbf{solid red line} represents the ROM mean prediction, the \textbf{dashed black line} indicates the true (test) trajectory, and the \textbf{shaded areas} denote the $\pm \sigma$ and $\pm 3\sigma$ confidence intervals.}
        \label{fig:stochastic-traj}
    \end{minipage}
\end{figure}

After training, the model can predict filtered trajectories based on the filtered initial condition. For each projection, the Variational Autoencoder (VAE) models a Gaussian distribution over the latent coordinates. The entropy of this distribution, which depends on the variance learned during training, reflects the projection's likelihood or uncertainty. At each time step, a latent vector is sampled to introduce uncertainty into the prediction. This uncertainty is then propagated through the autoregressive structure of the model. Consequently, each prediction incorporates the accumulated uncertainty from previous steps, as well as the uncertainty introduced by the current stochastic projection. We then decode these latent sample trajectories to project them back to the physical high-dimensional space. These thus constitute a family of plausible trajectories that are consistent with the underlying probability distribution of the flow. We refer to this family as an \textit{ensemble}, whose mean represents the most probable flow evolution and whose standard deviation represents the uncertainty. As expected, the uncertainty exhibits a growing trend as the dynamical roll-out progresses, as shown in Figure~\ref{fig:stochastic-traj}. The figure displays the \textit{mean} trajectory alongside confidence intervals defined by the standard deviation $\sigma$ of the ensemble. We also represent the true test trajectory to better evaluate the two statistical moments of the generated ensemble.

Figure~\ref{fig:stochastic-traj} also includes two snapshots of the velocity fields \(u\) and \(v\), taken 100 timesteps after the initial condition. Over time, due to the autoregressive nature of the model, uncertainty tends to accumulate, leading to increasing variance in the predictions. Despite this, we observe a strong alignment between the predicted mean and the true trajectory. Furthermore, before the uncertainty amplifies significantly, the confidence interval adapts to the evolving dynamics and transitions. This adaptation effectively balances the interval width with ground truth coverage. This relationship is formally quantified by the \textit{Prediction Interval Coverage Probability} (PICP), shown in Table~\ref{tab:metrics}. The PICP measures the fraction of true values that lie within the predicted confidence intervals, as
\[
\text{PICP} = \frac{1}{N} \sum_{i=1}^N \mathbf{1}\left(y_i \in [\mu_i - \sigma_i, \mu_i + \sigma_i]\right),
\]
where \( \mu_i \) and \( \sigma_i \) are the predicted mean and standard deviation, respectively. The empirical confidence intervals reported in Table \ref{tab:metrics} fall nearby those expected from a perfect Gaussian distribution, namely, \( 68\% \) within \( \pm\sigma \) and \( 99\% \) within \( \pm3\sigma \) which indicates that the model finds the mean and variance which distributes the error following a Gaussian-like behaviour. The intervals represent a practical trade-off between the ensemble spread and its ability to capture the true values.

This trade-off is further quantified by the \textit{Continuous Ranked Probability Score} (CRPS), also reported in Table~\ref{tab:metrics}. CRPS measures the accuracy of a probabilistic forecast, by
\[
\text{CRPS}(\mu, \sigma; x) = \sigma \left[ 
\frac{1}{\sqrt{\pi}} - 2\phi\left( \frac{x - \mu}{\sigma} \right) 
- \frac{x - \mu}{\sigma} \left( 2\Phi\left( \frac{x - \mu}{\sigma} \right) - 1 \right)
\right],
\]
where \( \phi \) and \( \Phi \) denote the standard normal probability density function (PDF) and cumulative distribution function (CDF), respectively. Lower CRPS values indicate more accurate probabilistic predictions, penalizing both bias and over, or under-confidence in the forecasted uncertainty. The low CRPS value reported here indicates a solid balance between interval conservatism and predictive accuracy. Table~\ref{tab:metrics} also includes the relative \( \ell_1 \) and \( \ell_2 \) errors, both highlighting the strong agreement between the ensemble mean and the true trajectories. These metrics are defined as
\[
\text{Rel-L1} = \frac{\sum |y_{\text{true}} - y_{\text{mean}}|}{\sum |y_{\text{true}}|} \times 100\%,
\quad
\text{Rel-L2} = \frac{\sqrt{\sum (y_{\text{true}} - y_{\text{mean}})^2}}{\sqrt{\sum y_{\text{true}}^2}} \times 100\%.
\]
\begin{table}[ht]
\centering
\begin{tabular}{l c}
\hline
\textbf{Metric} & \textbf{Value} \\
\hline
Relative L1 loss & 6.54\% \\
Relative L2 loss & 9.82\% \\
PICP$_{\pm 1\sigma}$ & 78.4\% \\
PICP$_{\pm 3\sigma}$ & 92.4\% \\
CRPS & 0.0567 \\
\hline
\end{tabular}
\caption{Evaluation metricss of the probabilistic forecasts.}
\label{tab:metrics}
\end{table}

\begin{figure}[H]
    \centering

    \begin{subfigure}[b]{0.33\textwidth}
        \centering
        \includegraphics[width=\linewidth]{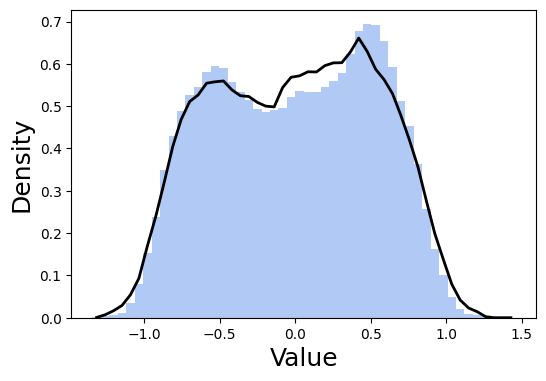}
        \caption{Comparison of the density of U and its prediction}
        \label{fig:pdf_u}
    \end{subfigure}
    \hspace{1em}
    \begin{subfigure}[b]{0.33\textwidth}
        \centering
        \includegraphics[width=\linewidth]{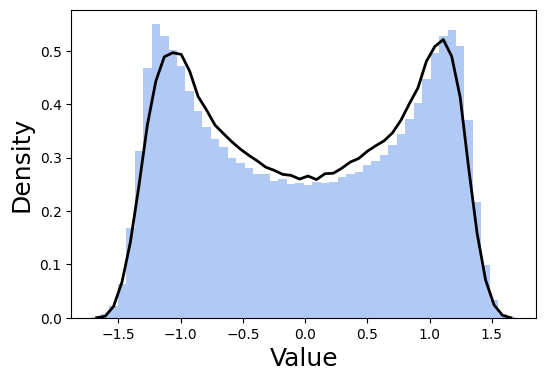}
        \caption{Comparison of the density of V and its prediction}
        \label{fig:pdf_v}
    \end{subfigure}

    \vspace{0.75em}

    \includegraphics[width=0.5\textwidth]{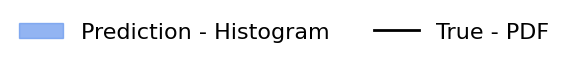}
    
    \caption{Probability density functions (PDFs) for fields U and V.}
    \label{fig:pdfs}
\end{figure}

Due to the chaotic nature of the system and the stochasticity inherent in the model, direct pointwise comparisons between individual snapshots and the ground truth are not always meaningful. Furthermore, our aim is to evaluate the prediction horizon beyond the true flow time window. We will therefore assess the model's long-term validity through statistical consistency. Specifically, we analyse whether the generated trajectory aligns well with the true dynamics in statistical terms, using probability density functions. We compute the probability density functions (PDFs) of the generated filtered velocity fields, and compare them to the true filtered PDFs. This frequentist approach ensures that, even if a trajectory deviates in its realization, it is still sampled from the same underlying statistical distribution, and that the long-term dynamical rollout remains consistent with that distribution. Figure~\ref{fig:pdfs}, illustrates this comparison of the PDFs, showing acceptable statistical agreement.

Finally, the Wasserstein distances are computed on the generated data and the true filtered trajectory, which are \( 0.0081 \) and \( 0.0269\) for \( U \) and \( V \), respectively. This indicates strong statistical agreement between the generated and reference distributions.
This confirms the conclusion that, for both fields \( U \) and \( V \), even if a particular trajectory deviates from the true one pointwise, or derives from a longer forecast roll-out, it still adheres to the underlying statistical distribution from which the true fields are drawn.
 
\section{Second Task: Gaussian process regression -- probabilistic small-scale closure}
\label{sec:Closure}
This second part describes the second identified task of the work: enhancing the fidelity of a reduced-order model acting on filtered flow fields by adopting a Gaussian process regression (GPR) framework. The advantage of Gaussian processes is that exact inference can be performed using matrix operations. It is also worth noting that the hyperparameters controlling the form of the Gaussian process are sparse and can be estimated directly from the data \citep{GPR}. This results in a lightweight model that enables fast training as well as quasi real-time closure and inference.  The objective is to establish a probabilistic mapping from low-dimensional filtered representations to full-resolution flow fields. 

We would like to emphasize that this method can be applied consecutively to virtually any low-fidelity data, whether generated by a machine learning model or numerical simulations such as URANS, or similar. The method introduced in Section~\ref{sec:ROM} is independent of the methodology presented here and merely serves as a necessary underlying dynamical model for low-fidelity data.

\subsection{Gaussian Process}

A Gaussian Process (GP) defines a distribution over functions, defined as
\[
f(x) \sim \mathcal{GP}(m(x), \mathcal{K}(x, x')),
\]
where \( m(x) \) is the mean function (typically centred to 0), and \( \mathcal{K}(x, x') \) is a positive-definite continuous covariance (kernel) function. Given training data \(\mathcal{D} = \{(x_i, y_i)\}_{i=1}^N\), where \(y_i = f(x_i) + \epsilon_i\), and \(\epsilon_i \sim \mathcal{N}(0, \sigma^2_\epsilon)\), we assume \(\bar{y}_i = 0\). The true underlying function \(f\) is unknown, and \(\sigma^2_\epsilon\) is a hyperparameter representing observation noise variance.

Let \(X = [x_1, \dots, x_N]^\top \in \mathbb{R}^{N \times d}\) be the input matrix, and \(Y = [y_1, \dots, y_N]^\top \in \mathbb{R}^{N \times 1}\) the corresponding output vector. The kernel matrix \(\mathcal{K}_{XX} \in \mathbb{R}^{N \times N}\) encodes pairwise similarities between training inputs, given
\begin{equation}\label{rbf}
\mathcal{K}(x_i,x_j|\tau) = \sigma^2 \exp\left(-\frac{1}{2l^2} \|x_i - x_j\|^2\right),
\end{equation}
where \(\tau = \{\sigma^2, l\}\) are the hyperparameters of the model; the output variance \(\sigma^2\) and the length-scale \(l\). The length-scale $l$ controls the smoothness of the function, The output variance $\sigma^2$ determines the typical scale of function values. The squared exponential (RBF) kernel is used here due to its smoothness, although for problems exhibiting rougher or periodic behavior, Matérn or periodic kernels might be more appropriate.

Let \(\mathcal{D}^* = \{(x_i^*, y_i^*)\}_{i=1}^M\) denote a set of unobserved test inputs, with \(X^* = [x_1^*, \dots, x_M^*]^\top \in \mathbb{R}^{M \times d}\). The key idea in Gaussian Processes is therefore that the training outputs \(Y\) and the latent function values at test inputs \(f^* = f(X^*)\) are jointly distributed as a multivariate Gaussian, as
\[
\begin{bmatrix}
Y \\
f^*
\end{bmatrix}
\sim \mathcal{N}\left(0,
\begin{bmatrix}
\hat{\mathcal{K}}_{XX} & \mathcal{K}_{XX^*} \\
\mathcal{K}_{X^*X} & \mathcal{K}_{X^*X^*}
\end{bmatrix}
\right),
\]
where \(\hat{\mathcal{K}}_{XX} = \mathcal{K}_{XX} + \sigma^2_\epsilon I_N\) includes the observation noise. This means that all data-points are partial observations drawn from the same Gaussian.

\begin{figure}[H]
\centering
\includegraphics[width=0.8\linewidth]{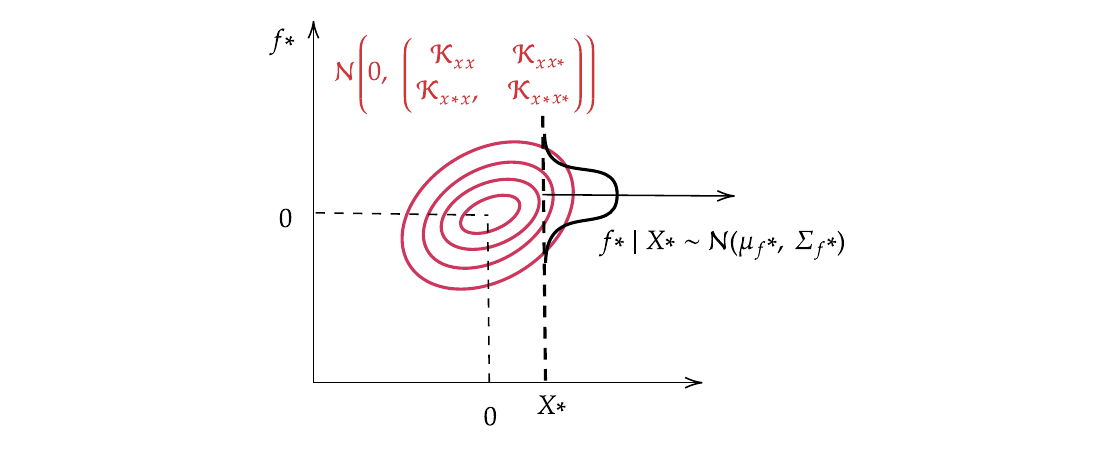}
\caption{Gaussian Process prediction example for a single training and test point (\(N=M=1\)).}
\label{fig:GP}
\end{figure}

Given partial observations of a joint Gaussian distribution, it is possible to compute the conditional distribution of unobserved variables, yielding a predictive distribution. The posterior predictive distribution at test inputs \(X^*\) is given by
\[
f^* \mid X^*, \mathcal{D} \sim \mathcal{N}(\mu_f, \Sigma_f),
\]
with:
\[
\mu_f = \mathcal{K}_{X^*X} \hat{\mathcal{K}}_{XX}^{-1} Y,
\quad
\Sigma_f = \mathcal{K}_{X^*X^*} - \mathcal{K}_{X^*X} \hat{\mathcal{K}}_{XX}^{-1} \mathcal{K}_{XX^*}.
\]
Figure~\ref{fig:GP} illustrates a 2D example where \(N = M = 1\).
The hyperparameters \(\tau\) and $\sigma_\epsilon$ are optimized by maximizing the log-marginal likelihood of the training outputs \(Y\) given inputs \(X\), as
\[
\log p(Y \mid X,\sigma^2_\epsilon, \tau) = \log \left[ \int p(Y \mid \tilde{f}, \sigma^2_\epsilon) p(\tilde{f} \mid X, \tau) d\tilde{f} \right]
\]
\[
 \log p(Y|X) = \log \left[ \int \mathcal{N}(Y|\tilde{f}, \sigma^2_\epsilon I) .\mathcal{N}(\tilde{f}|0, \mathcal{K}_{xx})d\tilde{f}\right]
\]
\[
 \log p(Y|X) = \log  \mathcal{N}(Y|0, \hat{\mathcal{K}}_{XX})
\]
where \[\hat{\mathcal{K}}_{XX} = \mathcal{K}_{XX} + \sigma^2_\epsilon I \]
\[
\log p(Y|X)  = -\frac{1}{2} Y^\top \hat{\mathcal{K}}_{XX}^{-1} Y 
                         - \frac{1}{2} \log \left| \hat{\mathcal{K}}_{XX} \right| 
                         - \frac{N}{2} \log(2\pi)
\]
This training process is summarized in Algorithm~\ref{alg:gpr-train}.
\begin{algorithm}[H]
\caption{Gaussian Process Hyperparameter Training}
\label{alg:gpr-train}
\begin{algorithmic}[1]
\State \textbf{Input:} Training dataset \(\mathcal{D} = \{(x_i, y_i)\}_{i=1}^N\)
\State initialise  hyperparameters \(\tau\), $\sigma^2_\epsilon$
\While{log-likelihood: \(\log p(Y \mid X, \tau, \sigma^2_\epsilon)\) not maximized}
    \State Compute covariance matrix \(\hat{\mathcal{K}}_{XX}\)
    \State Compute log-likelihood: \(\log p(Y \mid X, \tau, \sigma^2_\epsilon)\)
    \State Update hyperparameters \(\tau\), $\sigma^2_\epsilon$
\EndWhile
\end{algorithmic}
\end{algorithm}

At inference time, the predictive distribution for a new test point \(x^*\) is computed as shown in Algorithm~\ref{alg:gpr-infer}.

\begin{algorithm}[H]
\caption{Gaussian Process Inference}
\label{alg:gpr-infer}
\begin{algorithmic}[1]
\State \textbf{Input:} Training data \(\mathcal{D} = \{(x_i, y_i)\}_{i=1}^N\), test input \(x^*\), and trained hyperparameters \(\tau\)
\State Compute kernel submatrices:
\[
\mathcal{K} = 
\begin{bmatrix}
\hat{\mathcal{K}}_{XX} & \mathcal{K}_{XX^*} \\
\mathcal{K}_{X^*X} & \mathcal{K}_{X^*X^*}
\end{bmatrix}
\]
\State Compute posterior mean and variance:
\[
\mu_f = \mathcal{K}_{X^*X} \hat{\mathcal{K}}_{XX}^{-1} Y, \quad
\Sigma_f = \mathcal{K}_{X^*X^*} - \mathcal{K}_{X^*X} \hat{\mathcal{K}}_{XX}^{-1} \mathcal{K}_{XX^*}
\]
\State Predict output: sample \(y^* \sim \mathcal{N}(\mu_f, \Sigma_f)\)
\end{algorithmic}
\end{algorithm}

\subsection{GPR applied for closure}
\begin{figure}[H]
    \centering
    \includegraphics[width=1\linewidth]{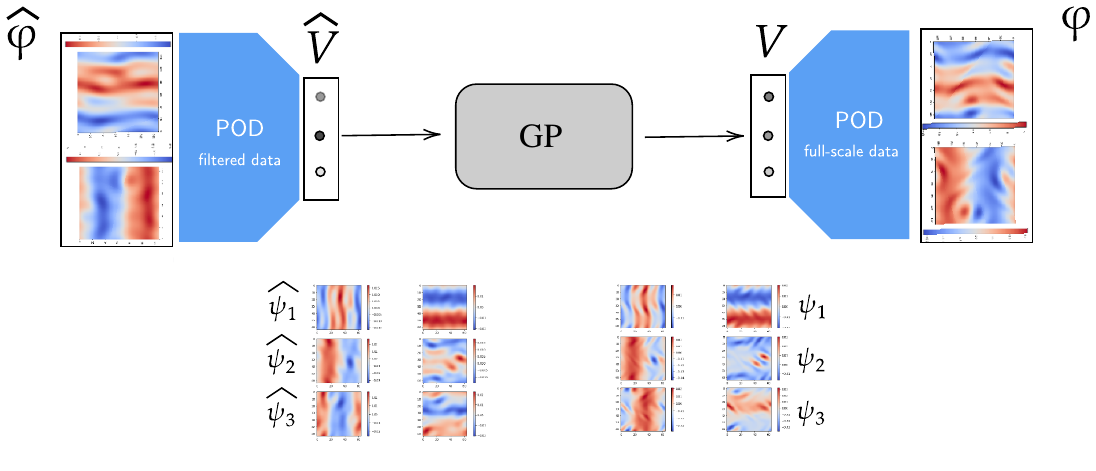}
    \caption{Mapping between low-pass filtered POD space and full-scale POD space}
    \label{fig:Projection}
\end{figure}

Rather than finding a mapping in physical space, which would result in too many degrees of freedom, we exploit the data's inherent low-dimensional structure and extract the mapping in a reduced space constructed using proper orthogonal decomposition (POD). POD is a data-driven, linear reduction technique that extracts the dominant coherent structures from the system. We apply POD to the snapshot matrix \( \varphi \in \mathbb{R}^{T \times D} \), where each column of \( \varphi \) represents the D-dimensional system state at a given time. Using Singular Value Decomposition (SVD), we write
\[
\varphi = U \Sigma V^T,
\]
where \( U \in \mathbb{R}^{D \times r} \) contains the spatial modes, \( \Sigma \in \mathbb{R}^{r \times r} \) is the diagonal matrix of singular values, and \( V \in \mathbb{R}^{T \times r} \) holds the temporal coefficients, with \( r \leq \min(T, D) \). We define \( \Psi = U \Sigma \), which combines the spatial structures with their energetic weights. A truncation at rank \( r = 3 \) is performed, since these encapsulate 98\% of the total energy in both the filtered and full-scale datasets. Better results could arguably be achieved by considering more modes, albeit at a higher training cost, without modifying the implementation. Finally, we use the symbol, \(^\wedge\), to indicate filtered data, while full-scale data is presented without any symbol. Figure \ref{fig:Projection} illustrates the mapping performed by the Gaussian Process Regression (GPR). 

Therefore, the system's state at time t can be approximated by a combination of the three identified POD modes, with
\[
\hat{\varphi}_t \approx a_t \hat{\psi_1} + b_t \hat{\psi_2} + c_t \hat{\psi_3}, 
\]
for the filtered and, with
\[
\varphi_t \approx \alpha_t \psi_1 + \beta_t \psi_2 + \gamma_t \psi_3,
\]
for the full-state data, respectively.

We aim to then learn a mapping, such that
\[
f: (a_t, b_t, c_t) \mapsto (\alpha_t, \beta_t, \gamma_t).
\]
Rather than learning a deterministic function, we place a prior over $f$:
\[
f \sim \mathcal{GP}(0, \mathcal{K}). 
\]
As a result, given the training data $\mathcal{D}$, we infer a posterior over functions
\[
p(f \mid \mathcal{D}, a, b, c).    
\]
The key advantage of Gaussian process regression (GPR) over classical regression methods is its inherently probabilistic formulation. By sampling from the predictive distribution \( y^* \sim f^* \), GPR produces an ensemble of outputs whose statistical properties are consistent with the underlying distribution of the flow. The associated variance naturally quantifies model confidence, providing not only pointwise predictions, but also credible intervals around them. This uncertainty-aware modelling approach aligns with recent turbulence modelling developments, which increasingly focus on learning statistical representations of flows rather than deterministic trajectories. It also aligns with the objective of this work, which is to remain within a statistical framework for reconstruction. 

\begin{figure}[H]
    \centering
    \includegraphics[width=1\linewidth]{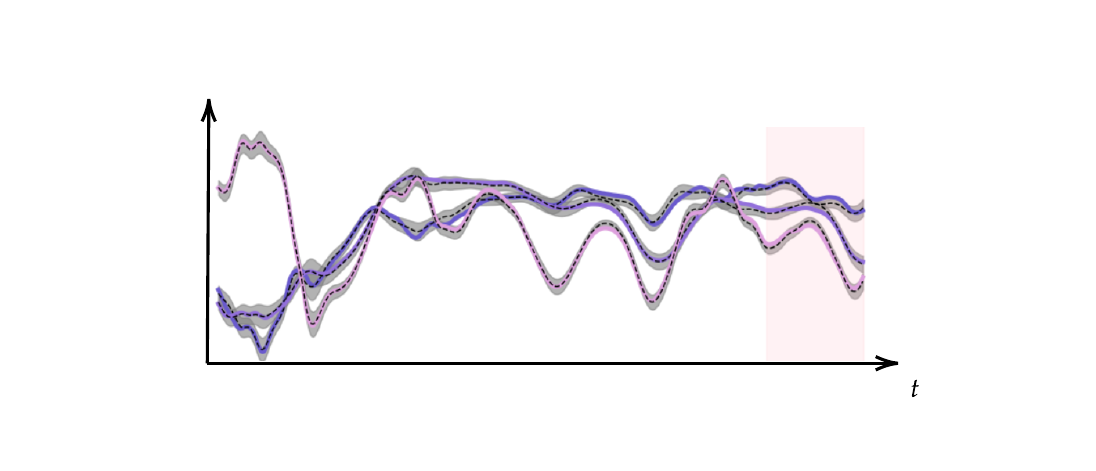}
    \includegraphics[width=1\linewidth]{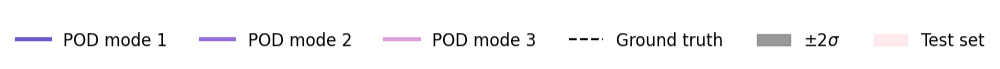}
    \caption{Gaussian Process Regression training and testing.}
    \label{fig:GP-training}
\end{figure}

Figure~\ref{fig:GP-training} shows the Gaussian process regression of the three reduced manifold variables in both the training and test sets.

The model performance is evaluated after the reverse projection to the high dimensional input space, using several metrics that quantify the accuracy of predictions and the quality of uncertainty estimates on the test set. We use the relative L1 error, the relative L2 error, the PICP introduced in \ref{ROM-perf} for a Confidence Interval (CI) of $\pm \sigma$ and $\pm 3\sigma$ and the CRPS. They are documented in Table \ref{tab:metrics_GP}.

\begin{table}[H]
\centering
\begin{tabular}{l c}
\hline
\textbf{Metric} & \textbf{Value} \\
\hline
Relative L1 loss      & 10.22\% \\
Relative L2 loss      & 11.52\% \\
PICP ($\pm 1\sigma$)  & 82.91\% \\
PICP ($\pm 3\sigma$)  & 100.00\% \\
CRPS                  & 0.0009 \\
\hline
\end{tabular}
\caption{Evaluation metrics on test set.}
\label{tab:metrics_GP}
\end{table}

\begin{figure}[H]
    \centering
    \includegraphics[width=0.8\linewidth]{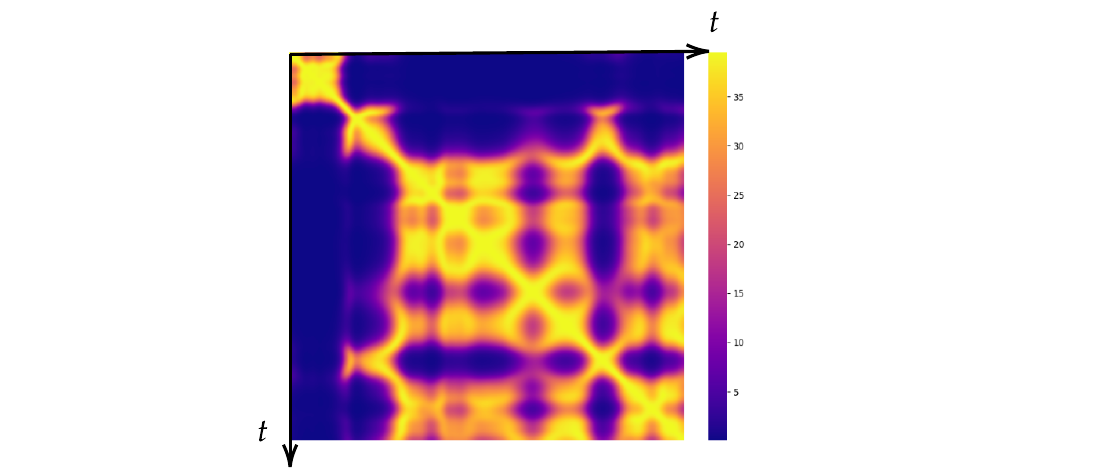}
    \caption{Training samples kernel \(\mathcal{K}_{XX}\)}
    \label{fig:Kernel}
\end{figure}

The performance of the Gaussian Process (GP) on the test set suggests that it acquired strong prior knowledge during training, enabling it to make meaningful predictions with reliable confidence intervals. The first-moment scores slightly above 10\% on the test set which indicates a decent level of fidelity to the test data. The Prediction Interval Coverage Probability (PICP) reaches 82\%, using intervals constructed at one standard deviations. The low Continuous Ranked Probability Score (CRPS) indicates that this coverage is achieved while maintaining minimal interval width. Following training, the model has identified the optimal set of three hyperparametrs, $\sigma^2$, $l$, and $\sigma_\epsilon^2$, which enables the constructions of the similarity kernel shown in Figure \ref{fig:Kernel}. This consequently allows inference on unseen data. The mapping can be instantaneously sampled from the GP as new data arrives, allowing for real-time closure.

The structure of the similarity kernel suggests that the snapshots are highly correlated with one another, even when they are temporally distant. This strong correlation indicates promising potential for generalization to unseen data.

\section{Full model prediction}

\begin{figure}[H]
    \centering
    \includegraphics[width=1\linewidth]{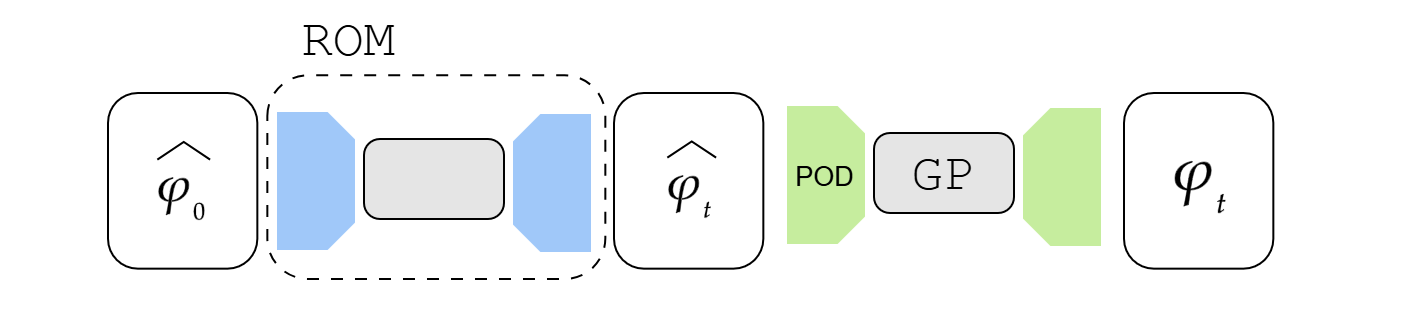}
    \caption{Data pipeline}
    \label{fig:Data Pipeline}
\end{figure}

As illustrated in the previous section, the Gaussian Process are trained to map filtered data to full-scale Direct Numerical Simulation (DNS) data. Initially trained on true filtered trajectories, the Gaussian process, when applied in the full model format, predicts full-scale data from the filtered trajectories generated by the dynamical reduced-order model (ROM) during inference. This establishes an end-to-end data pipeline where the input is the filtered initial condition and the output is the generated full-scale trajectories, as illustrated in figure \ref{fig:Data Pipeline}. 

\begin{figure}[htbp]
    \centering

    \begin{subfigure}[b]{0.32\textwidth}
        \includegraphics[width=\textwidth]{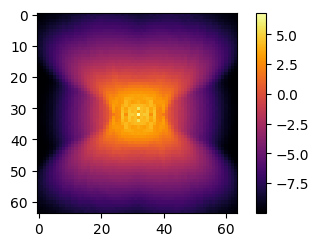}
        \caption{True flow}
        \label{fig:Truth}
    \end{subfigure}
    \hfill
    \begin{subfigure}[b]{0.32\textwidth}
        \includegraphics[width=\textwidth]{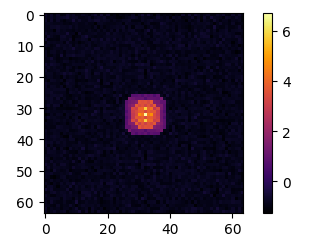}
        \caption{Flow predicted by the ROM}
        \label{fig:ROM}
    \end{subfigure}
    \hfill
    \begin{subfigure}[b]{0.32\textwidth}
        \includegraphics[width=\textwidth]{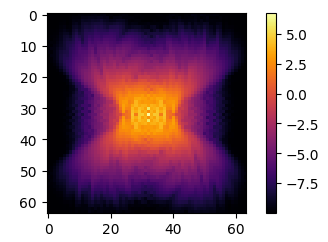}
        \caption{Flow predicted by the ROM \& GP}
        \label{fig:ROM_GP}
    \end{subfigure}

    \caption{Averaged energy spectrograms}
    \label{fig:reconstruction}
\end{figure}

When we condition the Gaussian Process (GP) model on new, unobserved inference data \( X^* \), which is generated by the Reduced Order Model (ROM), we obtain the posterior distribution:
\[
f^* \sim \mathcal{GP}(\mu_f, \mathcal{K}_{X X^*}),
\]
where \( \mu_f \) denotes the posterior mean, and \( \mathcal{K}_{X X^*} \) represents the updated covariance matrix. This matrix captures the similarity between the training data \( X \) and the inference points \( X^* \).  From this posterior, we can draw samples \(f^*\), which represent the predicted full-scale mode coefficients. The generated output successfully reconstructs filtered out frequencies as illustrated by the averaged spectrograms before and after the application of the Gaussian Process. Figure \ref{fig:reconstruction} shows these spectrograms for the kinetic energy. 

\begin{figure}[H]
    \centering

    \begin{subfigure}[b]{0.32\textwidth}
        \includegraphics[width=\textwidth]{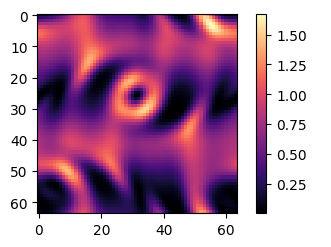}
        \caption{True flow}
        \label{fig:k_Truth}
    \end{subfigure}
    \hfill
    \begin{subfigure}[b]{0.32\textwidth}
        \includegraphics[width=\textwidth]{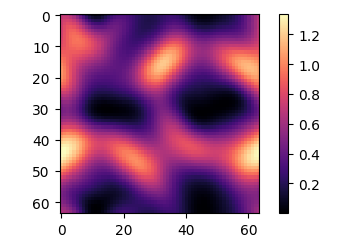}
        \caption{Flow predicted by the ROM}
        \label{fig:k_ROM}
    \end{subfigure}
    \hfill
    \begin{subfigure}[b]{0.32\textwidth}
        \includegraphics[width=\textwidth]{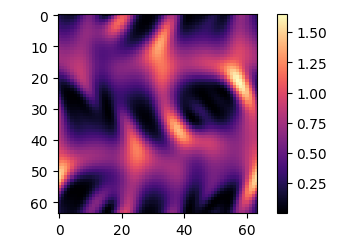}
        \caption{Flow predicted by the ROM \& GP}
        \label{fig:k_ROM_GP}
    \end{subfigure}

    \caption{Energy at a snapshot T = 120}
    \label{fig:reconstruction_k}
\end{figure}

The combination of the Reduced Order Model (ROM) and the Gaussian Process (GP) enables the generation of new trajectories that, while potentially deviating from the ground truth, remain statistically consistent with the underlying dynamics. At any given snapshot in time, the GP is capable of reconstructing missing or unresolved flow structures, such as small-scale eddies, as illustrated in Figure~\ref{fig:reconstruction_k}.

It is evident that one-to-one image comparisons with flows such as those shown in Figure \ref{fig:k_Truth} and \ref{fig:k_ROM_GP} appear different. Beyond a certain prediction horizon, pixel-to-pixel comparisons become irrelevant, since our framework does not simply learn and reproduce specific trajectories. Instead, it learns the underlying statistics of the flow and should not be expected to reconstruct identical realizations.

\subsection{Evaluation on long term predictions}
To ensure that our framework remains stable when making long-horizon predictions, we assess whether the statistical properties of the generated flow are preserved over time. Specifically, we compare the probability density functions (PDFs) of the predicted full-scale flow with those of the validation set.

\begin{figure}[htbp]
    \centering
    \includegraphics[width=0.8\linewidth]{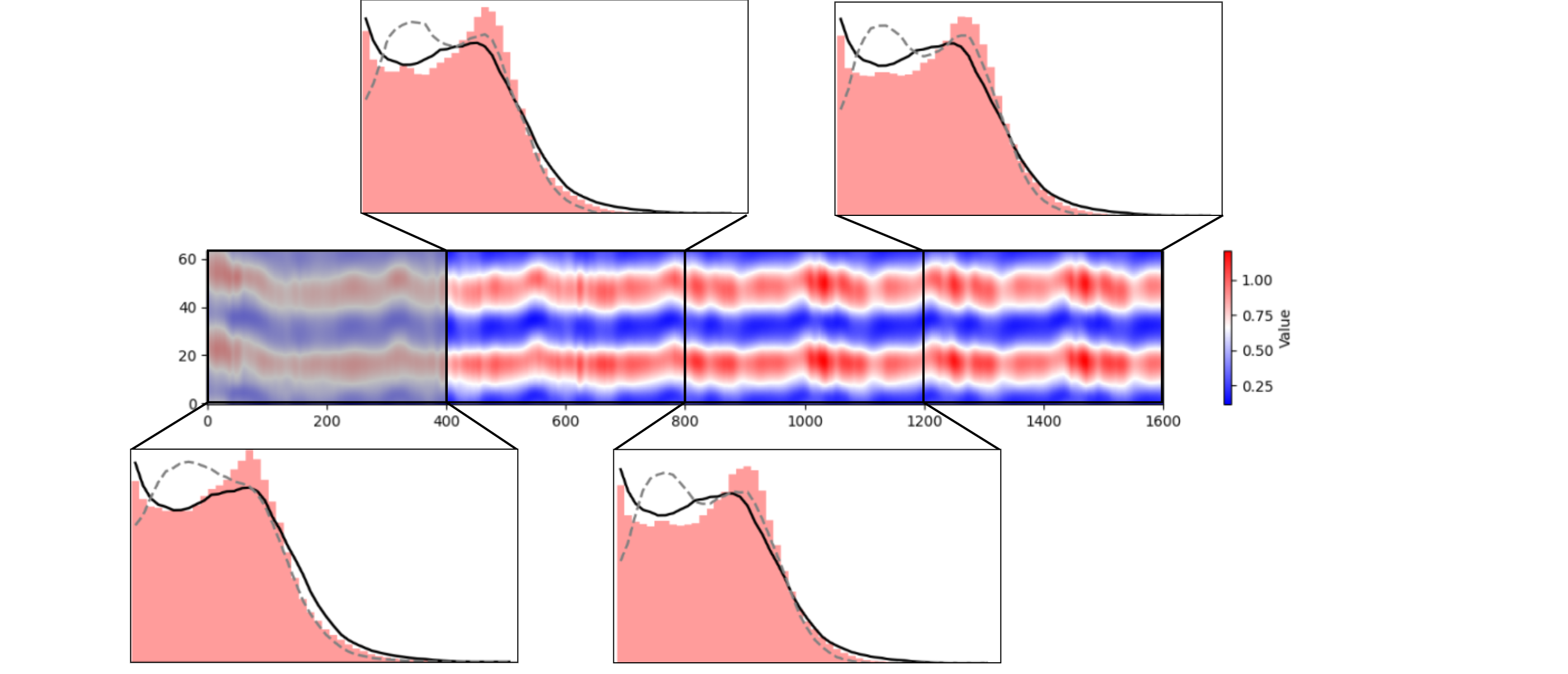}
    \includegraphics[width=1\linewidth]{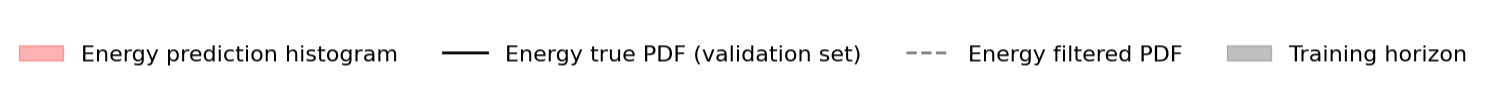}
    \caption{Kinetic energy over a dynamical rollout of 1600 snapshots. The data distribution for each 400-snapshot interval is compared with the PDF of the validation set.}
    \label{fig:longrollout}
\end{figure}

The validation set consists of 500 snapshots, while the ROM+GP framework generates a trajectory of 1600 snapshots.  Figure~\ref{fig:longrollout} displays the kinetic energy roll-out computed on a downsampled \(64 \times 64\) grid (reduced to 64 values via spatial averaging). We compare the histogram of each 400 snapshots slice alongside the reference PDF from the validation set. We recall that both the ROM and the GP were trained on the first 400 snapshots of the filtered DNS or full-resolution DNS, respectively. Figure ~\ref{fig:longrollout} also displays the energy density before the Gaussian Process closure.

We observe that even when predicting over a time horizon four times longer than the training window, the closure fixes the filtered rollout energy statistics and preserves it in forecast mode without significant drift. To quantify this observation, we compute the Wasserstein distance between the predicted energy density and the validation set distribution, using windows of 200 snapshots.

\begin{figure}[H]
    \centering
    \includegraphics[width=0.6\linewidth]{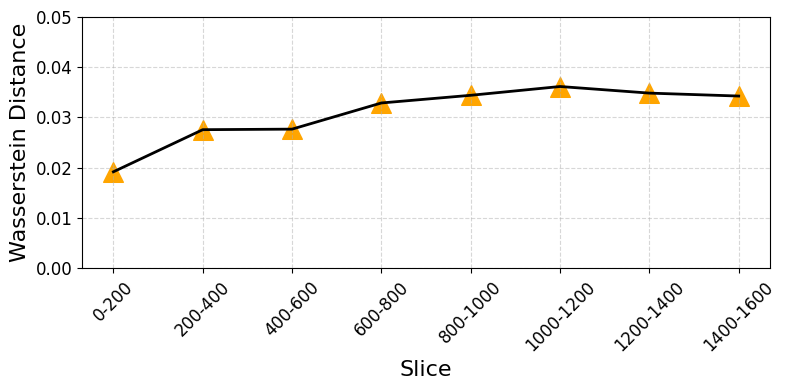}
    \caption{Wasserstein distance between the energy density of the validation set and that of the predicted flow, computed over consecutive 200-snapshot intervals.}
    \label{fig:placeholder}
\end{figure}
This result demonstrates that the statistical properties of the flow remain stable throughout the forecasting window. Despite evolving over a time horizon significantly longer than the training window, the model maintains both spatial and temporal complexity without substantial statistical degradation.

\subsection{Comparison to alternative ML architectures}
\label{sub-sec:Diff_VAE}

We compare the performance of our GPR-based model against state-of-the-art probabilistic machine learning predictors, namely the Variational Autoencoder (VAE) and Diffusion models. This choice is motivated by the remarkable capabilities of these statistical distribution emulators, particularly in image generation and hyper-resolution tasks. In this context the multi-scale closure task is performed using these ML architectures. Appendix~\ref{subsec:alternative_arch} provides the implementation details for each architecture, along with further justification for their selection as benchmarks.

The results for both the VAE and the diffusion model are summarized over the test set in Table~\ref{tab:comparison}. Both were trained on the same flow resolution as the GP and with identical train and test sets.  

\begin{table}[ht]
\centering
\begin{tabular}{l c c c}
\hline
\textbf{Metric} & \textbf{GP} & \textbf{VAE} & \textbf{Diffusion Model} \\
\hline
Relative L1 loss (\%)     & \textbf{10.22}  & 20.34 & 39.56 \\
Relative L2 loss (\%)     & \textbf{11.52}  & 22.77 & 41.96 \\
PICP ($\pm 1\sigma$) (\%)  & \textbf{81.94} & 19.34 & 50.39 \\
PICP ($\pm 3\sigma$) (\%)  & \textbf{100.00} & 52.22 & 89.25 \\
CRPS                  & \textbf{0.0009} & 0.1127 & 0.1915 \\
\hline
\end{tabular}
\caption{Comparison of evaluation metrics between the GP method, VAE, and diffusion model.}
\label{tab:comparison}
\end{table}
The Gaussian process outperforms these two baselines for both the first and second moment metrics on test set.
It should be noted, however, that although the diffusion model demonstrates state-of-the-art performance in image generation and does not rely on any prior assumptions about the data structure, its training and inference costs are notably high. Specifically, generating a single image requires numerous denoising steps (1000 in our implementation), meaning the model must perform a thousand inference passes to produce a unique sample. This results in a significant computational expense. We hypothesise that, with a larger model and a more exhaustive search for optimal hyperparameters, diffusion could perform better than shown here, especially with access to more data. However, such a search would incur prohibitive computational costs.
The VAE needs to perform inference once per sample, resulting in faster predictions but potentially high computational cost, especially with deeper encoders and decoders. The GP, in contrast, generates all samples at once, directly in the reduced manifold space by treating different snapshots as partial observations of a single underlying Gaussian distribution. Notably, only three hyperparametrs were optimized during training, leading to a considerably more efficient and interpretable model.

\section{Conclusion}
\label{sec:conclusion}
This study introduces a predictive modeling framework for the long-term generation and forecasting of chaotic trajectories representative of turbulent regimes. The proposed approach adopts a fully stochastic formulation that exploits scale separation to reproduce the statistical properties of the original system. In this framework, the coherent large-scale dynamics, identified through data filtering, are first predicted, while the influence of small-scale fluctuations is incorporated through a stochastic representation. This simplification enables the reduced-order model (ROM) to focus on the evolution of time-dependent dynamics rather than the detailed reconstruction of fine-scale features. Once the filtered trajectories are inferred, a secondary, time-independent model probabilistically maps these results back to the full-scale space. Notably, the two models operate independently and are designed to be modular: the ROM can emulate large-scale dynamics while being coupled with alternative closure schemes, and the small-scale learning strategy can serve as a closure mechanism for alternative low-fidelity dynamical models. This plug-and-play architecture provides enhanced flexibility and adaptability across a broad range of turbulent flow applications.

The reduced-order model (ROM) used in this framework is based on our recent work, \textit{UPdROM}~\citep{UPdROM}. Although it was originally trained and validated on the deterministic flow past a blunt object, the model's stochastic nature enables it to generate ensembles of trajectories whose statistics closely align with the Kolmogorov flow distribution. This distribution is considered here as a benchmark test case. The model learns a probabilistic mapping to a low-dimensional manifold using a variational autoencoder (VAE). Dynamics within this latent space are captured by a Transformer architecture. Furthermore, by sampling stochastically from the VAE at each time step, the model generates ensembles of trajectories. The ensemble mean remains faithful to the true data, while the variance demonstrates strong coverage properties, as confirmed by the Prediction Interval Coverage Probability (PICP). The trade-off between ensemble spread and coverage is quantified by a Continuous Ranked Probability Score (CRPS) approaching zero. To close the spatial structures, we use a Gaussian process (GP) to map from the filtered space to the full-scale space within their respective POD subspaces. Using only three POD modes, we can faithfully reconstruct the multi-scale data, as demonstrated by strong test-set performance. This includes accurate first-moment statistics (with low $\ell_1$ and $\ell_2$ errors) and robust second-moment statistics (as evidenced by the PICP and CRPS).

Building upon this scale-separated stochastic architecture, we demonstrate a predictive model that maintains statistical consistency and predictive accuracy over extended forecasting horizons. The efficiency of the proposed stochastic prediction framework opens new possibilities for real-time training, fine-tuning, and inference, particularly in the context of flow control applications.

We benchmarked these results against those of other probabilistic mapping baselines, specifically a variational autoencoder (VAE) and a denoising diffusion model, using their standard implementations. The GP outperformed both methods across all metrics considered. This is particularly surprising given that diffusion models are often considered the gold standard in the community for tasks such as image generation and super-resolution, which are closely related to our problem. However, a standard denoising diffusion algorithm infers a single snapshot by recursively denoising a white noise image and typically requires a thousand denoising steps per snapshot. This means that, to generate one clean sample, the model must be inferred a thousand times. In contrast, VAE requires only one inference per snapshot, making it more cost-effective. However, GP generates the entire dynamic rollout with a single inference since all snapshots are treated as partial observations from the same multivariate Gaussian distribution. Furthermore, this distribution is parametrized by just three optimized parameters, ensuring fast training and significantly faster inference.

\paragraph{Funding Statement}
This research was supported by grants from Sound.AI program, an international AI training initiative, co-funded by the \textit{European Union} under the \textit{Marie Skłodowska-Curie program}. 

\begin{figure}[ht]
\centering
\includegraphics[width=0.20\textwidth]{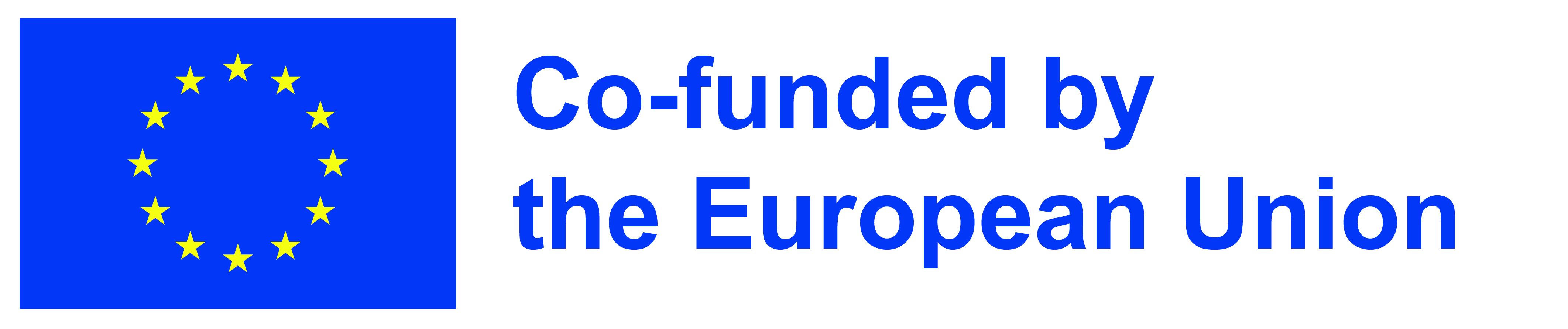}\hspace{1cm}
\includegraphics[width=0.17\textwidth]{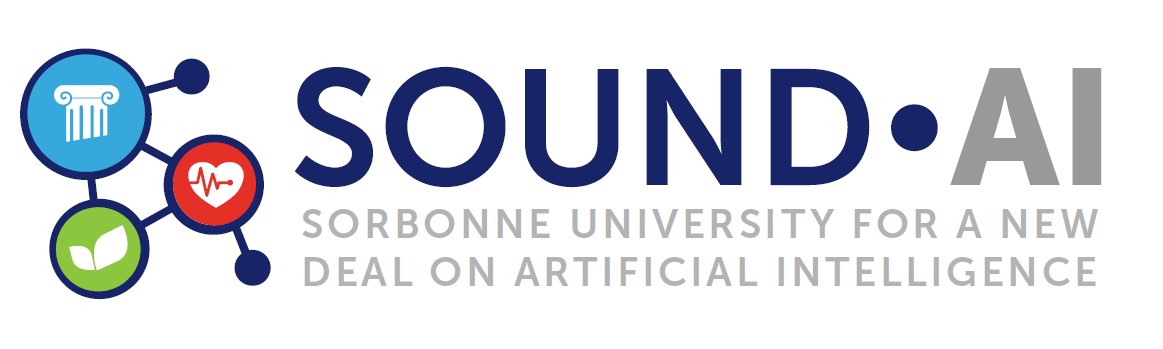}\hspace{1cm}
\includegraphics[width=0.12\textwidth]{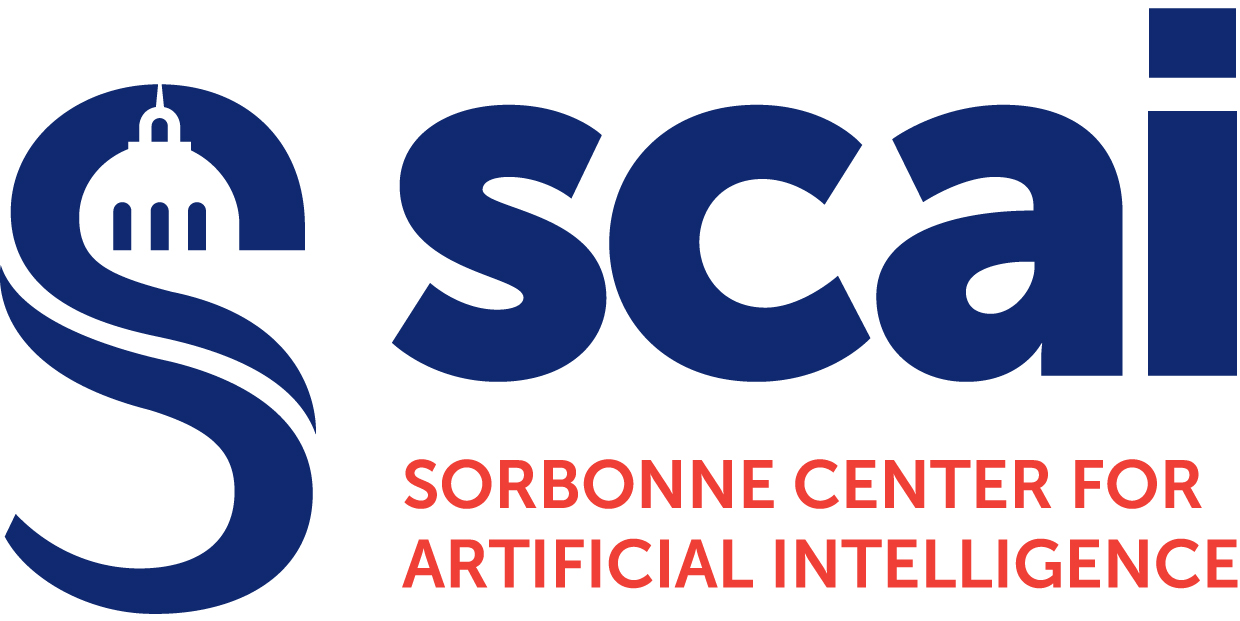}\hspace{1cm}
\includegraphics[width=0.15\textwidth]{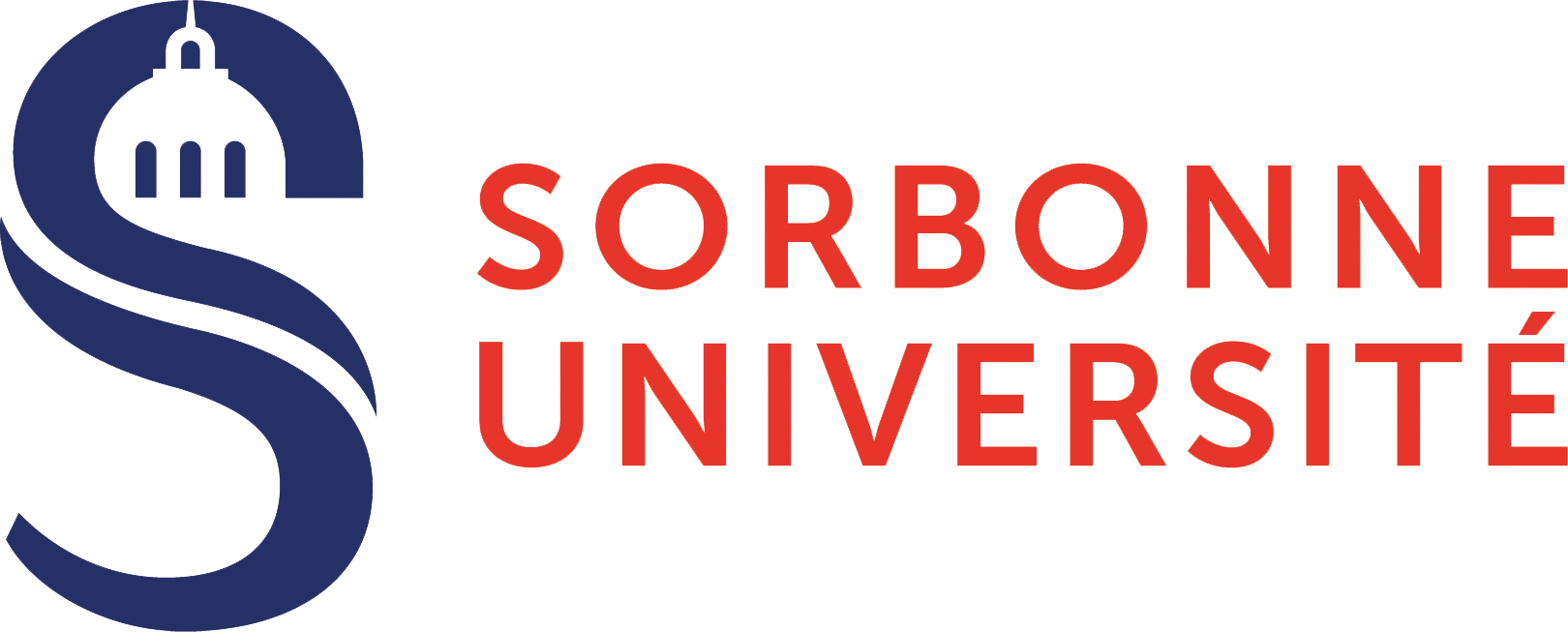}\hspace{1cm}
\end{figure}

\paragraph{Competing Interests}
None

\paragraph{Data Availability Statement}
The Reduced Order Model (ROM) used for the first task (Section \ref{sec:ROM}) is openly available on GitHub \citep{GitHubRepo}. A GitHub repository containing the code and access to the experimental data used throughout the entire modelling pipeline is available in \citep{Gitrepo_scale_separation}.

\paragraph{Ethical Standards}
The research meets all ethical guidelines, including adherence to the legal requirements of the study country.

\paragraph{Author Contributions}
\textbf{Ismaël Zighed } : Conceptualization, Methodology, Data curation, Data Visualisation, Formal analysis, Validation, Writing original draft. \textbf{Nicolas Thome} : Resources, Supervision, Validation, Writing – review \& editing.  \textbf{Patrick Gallinari} : Resources, Supervision, Validation, Writing – review \& editing. \textbf{Taraneh Sayadi} :  Conceptualization, Methodology, Resources, Formal analysis, Supervision, Validation, Writing – review \& editing. \\
All authors approved the final submitted draft.

\bibliographystyle{plainnat}
\bibliography{biblio}

\begin{appendix}
    \section{Appendix}
\label{sec:appendix}
\subsection{Filtered dynamics - ROM architecture}
\label{subsec:ROMarch}
\begin{table}[h!]
\centering
\begin{tabular}{ll}
\hline
\textbf{Parameter} & \textbf{Value} \\
\hline
Number of trajectories & 8 \\
Time dimension & 400 \\
Spatial dimensions & $64 \times 64$ \\
Flows & $u, v$ \\
Input dimension & $64 \times 64 \times 2 = 8192$ \\
Latent dimensions & 64 \\
Kullback-Leibler regularization & $5 \times 10^{-4}$ \\
Attention blocks & 2 \\
Attention heads & 16 \\
Prediction horizon & 30 \\
Lookback window & 30 \\
\hline
\end{tabular}
\caption{Model and data configuration parameters.}
\end{table}

\subsection{Full-scale closure - alternative architectures}
\label{subsec:alternative_arch}
The diffusion model used as a benchmark in \ref{sub-sec:Diff_VAE} is inspired by the original Denoising Diffusion Probabilistic Models (DDPM) framework \citep{DDPM}, where the model is trained to iteratively denoise data corrupted by Gaussian noise. The denoising network employs a U-Net style auto-encoder, using max-pooling layers and convolutional kernels of size 3 to progressively downsample input images from \(64 \times 64\) and the two input channels $(u,v)$ to a latent representation of size \(4 \times 4\) with 256 feature channels.

Formally, the forward noising process is defined by a noise schedule \(\{\beta_t\}_{t=1}^T\), with \(T=1000\) noising and denoising steps, gradually adding noise to the clean data \(x_0\) to produce noisy samples \(x_t\) via

\[
q(x_t \mid x_0) = \mathcal{N}\left(x_t; \sqrt{\bar{\alpha}_t} x_0, (1 - \bar{\alpha}_t) \mathbf{I} \right),
\]

where \(\bar{\alpha}_t = \prod_{s=1}^t (1 - \beta_s)\). The model is trained to predict the noise \(\epsilon\) added at each step by minimizing the mean squared error loss:

\[
\mathcal{L} = \mathbb{E}_{x_0, \epsilon, t} \left[ \|\epsilon - \epsilon_\theta(x_t, t)\|^2 \right],
\]

where \(\epsilon_\theta\) is the noise predicted by the network at timestep \(t\). The inference is performed by sampling from the learned reverse process starting from Gaussian noise conditioned on the filtered data, reconstructing clean samples through iterative denoising. Training utilizes the Adam optimizer over 2000 epochs. Conceptually, the diffusion model learns a parameterization of the reverse conditional Gaussian transitions that gradually remove noise. These Gaussian transitions are defined directly in the original high-dimensional input space. \\

The Variational Auto-Encoder (VAE) used as the other benchmark in \ref{sub-sec:Diff_VAE} is motivated by the assumption that the data distribution lies on a low-dimensional manifold, as is often the case in dynamical systems applications. The VAE also aims to approximate the posterior distribution of the latent variables given the observed data \( x_{\text{data}} \) by learning parameters \((\mu_\theta, \Sigma_\theta)\) of a Gaussian distribution in a latent space of significantly lower dimensionality than the original input space \citep{VAE}, unlike Diffusion models. This low-dimensional representation facilitates much more efficient inference and captures the essential structure of the data. It does however rely on strong hypothesis that such low-dimensional projection exists. 
The VAE employed in this work is a fully connected neural network that gradually compresses the input data from dimension \(64 \times 64\) down to a 4-dimensional latent space. The model is trained by optimizing the variational lower bound, expressed as

\[
\mathcal{L} = \mathrm{KLD}\big(q_\theta(z|X) \,\|\, p(z)\big) - \mathbb{E}_{z \sim q_\theta} \big[ \log p_\theta(X|z) \big],
\]

where \(q_\theta(z|X)\) is the approximate posterior, \(p(z)\) the prior, and \(p_\theta(X|z)\) the likelihood. In practice, we introduce a weighting factor on the Kullback-Leibler divergence term to prioritize accurate reconstruction over strict adherence to Gaussianity in the latent space, thus balancing the trade-off between fidelity and regularization. The weighting factor used in this example is $0.01$.

\end{appendix}

\end{document}